\documentclass[prb,twocolumn,superscriptaddress,amsmath,amssymb,floatfix,preprintnumbers]{revtex4}
\usepackage{amsmath}
\usepackage{graphicx}
\usepackage{dcolumn}
\usepackage{color}
\usepackage{lipsum}
\DeclareGraphicsExtensions{.png .jpg .pdf}

\usepackage{dsfont}

\begin{document}

\title{Electronic and transport properties of anisotropic semiconductor quantum wires}

\author{S. M. Cunha}\email{sofiacunha@fisica.ufc.br}
\author{D. R. da Costa}\email{diego_rabelo@fisica.ufc.br}
\author{L. C. Felix}
\author{Andrey Chaves}
\author{J. Milton Pereira Jr.}\email{pereira@fisica.ufc.br}
\affiliation{Departamento de F\'isica, Universidade Federal do Cear\'a, Campus do Pici, Fortaleza, Cear\'a, Brazil}

\date{ \today }

\begin{abstract}

Within the effective-mass approximation, we theoretically investigated the electronic and transport properties of 2D semiconductor quantum wires (QWs) with anisotropic effective masses and different orientations with respect to the anisotropic axis. The energy levels in the absence and presence of an external magnetic field are analytically calculated, showing: (i) a strong dependence on the spacing of energy levels related to the alignment QW angle and the anisotropy axis; and (ii) for non-null magnetic field, the quantum Hall edge states are significantly affected by the edge orientation. Moreover, by means of the split-operator technique, we analyzed the time evolution of wavepackets in straight and V-shaped anisotropic QWs and compared the transmission probabilities with those of isotropic systems. In the anisotropic case we found damped oscillations in the average values of velocity in both $x$ and $y$ directions for a symmetric Gaussian wavepacket propagating along a straight wide QW, with the oscillation being more evident as the non-collinearity between the group velocity and momentum vectors increases.
\end{abstract}

\maketitle

\section{Introduction} 

In the last two decades, the production of graphene has led to a significant level of interest on the physics of layered materials [\onlinecite{graphene1}-\onlinecite{layered6}]. This interest is not only due to its possible future technological applications, but also because it provides the possibility to probe interesting phenomena predicted by quantum field theories not found in conventional semiconductors and metals. Along with the investigation of basic properties of these materials, there has also been a significant effort to develop devices that can benefit from their two-dimensional ($2$D) character. In that respect, the introduction of additional confinement by creating $1$D (quantum wires (QWs)) and $0$D (quantum dots) structures becomes relevant [\onlinecite{continuum0}-\onlinecite{confine7}], since these are known to modify the electronic spectra and the transport properties of the structure in comparison with the pristine sample.

Most recently, there is a growing interest in single layers of black phosphorus (BP), also known as phosphorene [\onlinecite{continuum}-\onlinecite{phosphorene5}] which is a semiconductor with a puckered structure, due to $sp^3$ hybridization and displays a tunable bandgap [\onlinecite{phosphorene1},\onlinecite{phosphorene2}]. In addition, phosphorene presents a highly anisotropic band structure and thus an anisotropic effective mass [\onlinecite{phosphorene6, continuum0, continuum, continuum1, phosphorene1, phosphorene2, phosphorene3, phosphorene4, phosphorene5}]. Another material that has attracted attention due to its anisotropic properties is single layer Arsenic (arsenene) [\onlinecite{arsenene1}-\onlinecite{arsenene6}], a semiconductor also with a puckered structure. Due to the highly anisotropic band structures of such crystals, their electrical conductivity, thermal conductivity and optical responses are found to be strikingly dependent on the crystallographic directions [\onlinecite{phosphorene3},\onlinecite{phosphorene4},\onlinecite{anisotropic1}-\onlinecite{anisotropic9}]. In particular, one possible consequence of the anisotropy may be seen in the electronic confinement caused by the presence of constraints such as external gates or crystal terminations. In that case, a dependence of the confined states on the direction of the alignment of the constraint may arise.

In the present paper we investigate the electronic and transport properties of anisotropic materials in which a 1D confining potential has been imposed. The work proceeds as follows: Initially we investigate the case of 1D confinement in an anisotropic system (i.e. a QW) in which the QW orientation may not match the anisotropy axis of the sample. In order to do that, we employ an effective mass model in which the anisotropy is encoded in the direction-dependent effective mass. Next, we show results for the spectra of confined states for different orientation angles of QW edges in the presence of an external magnetic field. By using the split-operator technique [\onlinecite{SplitOperator, SplitOperator0, SplitOperator1, SplitOperator3, SplitOperator4, SplitOperator6, SplitOperator7, SplitOperator8, SplitOperator9, SplitOperator10, SplitOperator11, SplitOperator12}], we then present results for the time-evolution of a Gaussian wavepacket propagating in an anisotropic QW that presents a ``bend'', i.e. the orientation of the QW with regards to the anisotropy axes changes along the longitudinal direction. We numerically investigate the electronic scattering of the propagated wavepacket at the bend caused by the mismatch between the electronic subbands at each QW region, which is an evidence of their dependence on the orientation angle. In addition, we calculate the average velocity values ​​for the $x$ and $y$ directions of an initially symmetrical Gaussian wavepacket propagating along a large QW in order to analyze the non-specular reflections at the QW edges and the combination of effects due to the anisotropy and system geometry. 

The paper is organized as follows: in Sec.~\ref{sec.classic} we present the analytical model for anisotropic classic systems taking as starting point an effective mass model. We show the spectrum of confinaded states for QWs anisotropic systems with different orientation angles with and without an external magnetic field in Sec.~\ref{sec.wires}. The influence of an anisotropic QW formed by leads with  different alignment angles in the scattering initial Gaussian wavepacket is studied in Sec.~\ref{sec.scattering}. Our conclusions are presented in Sec.~\ref{sec.conclusions}.

\section{Anisotropic classic systems}\label{sec.classic}

\begin{table}[b]
\centering
\caption{Electron effective masses in the $x$ and $y$ directions for phosphorene and arsenene in units of free electron mass ($m_0$).[\onlinecite{continuum1}]}
\resizebox{\linewidth}{!}{
\begin{tabular}{p{2.5cm}p{2.5cm}p{2.5cm}}
\hline \multicolumn{1}{l}{} &  phosphorene  & arsenene\\ \hline
$m_{x}/m_0$          &   1.01           & $0.23$ \\
$m_{y}/m_0$          &   0.19            & $1.22$ \\ \hline
\end{tabular}%
}\label{table}
\end{table} 

Let us consider an anisotropic 2D system in which the anisotropy is introduced as direction-dependent effective masses. Among an extensive list of anisotropic materials, such as BP [\onlinecite{continuum}-\onlinecite{phosphorene5}], arsenene [\onlinecite{arsenene1}-\onlinecite{arsenene6}], ReS$_2$ [\onlinecite{ReS2}], TiS$_3$ [\onlinecite{TiS3}], and others, the first two are the most prominent ones, and for that reason why, we henceforth assume parameters suitable for these materials. Similar qualitative results discussed along this work are expected for any of the above mentioned anisotropic materials. Effective mass models have been shown to give a reasonable description of the low-energy spectrum of phosphorene and arsenene.[\onlinecite{andreyStarkEffect, gabriel}] In general, in the theoretical analysis of such system, it is convenient to chose coordinate axes in such a way that they match the anisotropy directions (henceforth known as the $x$ and $y$ directions, with $m_x$ and $m_y$ being the effective masses along each direction, respectively). Table I presents the values of electron effective masses for both phosphorene and arsenene. However, as shown below, it is necessary in the present case to consider a more general configuration. Thus, in general the Hamiltonian is given by
\begin{eqnarray}\label{eq:1}
H = \frac{p^2_x}{2m_x}+\frac{p^2_y}{2m_y}.
\end{eqnarray}
A curve of constant energy in momentum space is then an ellipse. A more complicated but also more interesting case is when the coordinate axes are not parallel to the anisotropy axes. We can obtain that by rotating the coordinate system in momentum space, such that the semi-major axis of the elliptical constant energy curve is rotated by an angle $\alpha$ around the $z$ axis. That give us: $p_x = p'_x\cos\alpha - p'_y\sin\alpha$  and $p_y = p'_x\sin\alpha + p'_y\cos\alpha$, where the primed terms correspond to the new, rotated coordinate system. Thus, we can now obtain the Hamiltonian as
\begin{eqnarray}\label{eq:3}
H = \frac{p'^2_x}{2\mu_1} + \frac{p'^2_y}{2\mu_2} + \frac{p'_x p'_y}{\mu_3},
\end{eqnarray}
with 
\begin{subequations}
\begin{eqnarray}
\frac{1}{\mu_1} &=& \frac{\cos^2\alpha}{m_x} + \frac{\sin^2\alpha}{m_y},\\
\frac{1}{\mu_2} &=& \frac{\sin^2\alpha}{m_x} + \frac{\cos^2\alpha}{m_y},\\
\frac{1}{\mu_3} &=& \left(\frac{1}{m_y} - \frac{1}{m_x}\right)\sin\alpha\cos\alpha.
\end{eqnarray}
\end{subequations}

From Eq.~(\ref{eq:3}) we find
\begin{eqnarray}\label{eq:4}
p'_y = \pm \sqrt{2\mu_2 E - \left(\frac{\mu_2}{\mu_1} - \frac{\mu_2 ^2}{\mu_3 ^2}\right)p'^2_x} - \frac{\mu_2}{\mu_3}p'_x.
\end{eqnarray}
It can be immediately seen that for $m_x = m_y$ (i.e. $1/\mu_3 = 0$) we obtain $p'_y = \sqrt{2\mu E - p^{'2}_x}$, with $\mu_1 = \mu_2 = \mu$, as expected for the isotropic case. Let us now obtain the components of the velocity vector. An important feature of anisotropic systems is the fact that the velocity is usually not collinear with the momentum vector, as shown below by computing $v'_i = \partial E/\partial p'_i$ for $i=x$ and $y$. Thus, the velocity components are given by
\begin{eqnarray}\label{eq:5.1}
v'_x = \frac{p'_x}{\mu_1} + \frac{p'_y}{\mu_3} \ \ , \ \ v'_y = \frac{p'_y}{\mu_2} + \frac{p'_x}{\mu_3},
\end{eqnarray} 
where it is seen that $v'_x$ ($v'_y$) can be non-zero even if $p'_x$ ($p'_y$) vanishes. 

\begin{figure}[!b]
\centerline{\includegraphics[width = 0.75\linewidth]{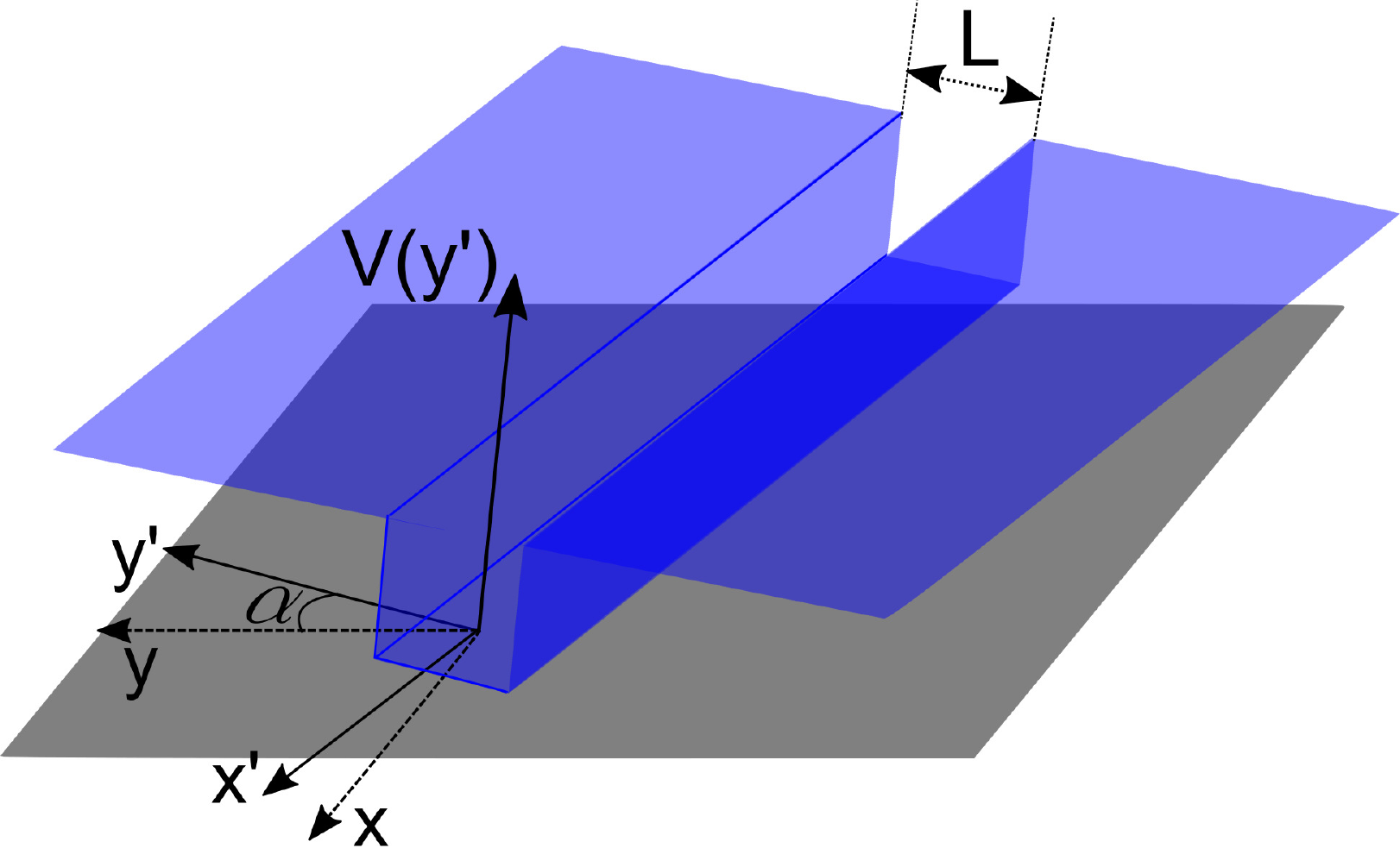}}
\caption{(Color online) Schematic representation of the rotated QW defined electrostatically by the $1$D square-well potential $V(y') = V_0\left[\Theta\left(-y'\right)+\Theta\left(y'-L\right)\right]$ with width $L$ and $V_0>0$. $\alpha$ is the rotation angle with respect to the crystallographic directions ($x$ and $y$), defining the new primed coordinates ($x'$ and $y'$).} \label{fig:anisotropic}
\end{figure}

\section{Anisotropic QWs}\label{sec.wires}
\subsection{In the absence of magnetic field}\label{sec.B.eq.0}

\begin{figure}[!t]
\centerline{\includegraphics[width = \linewidth]{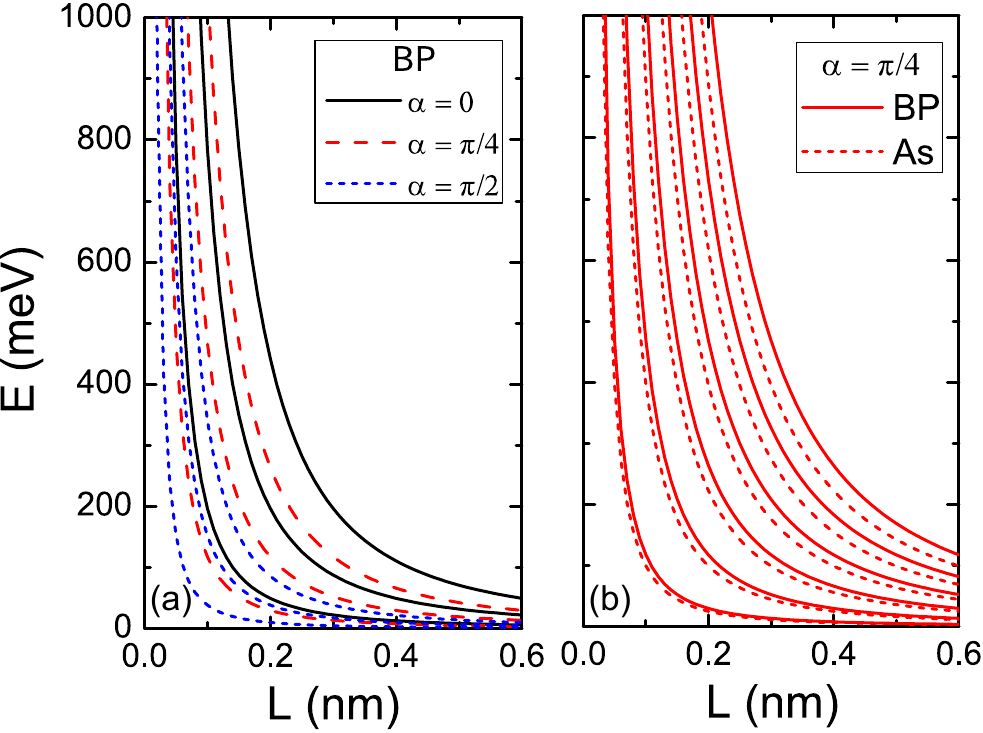}}
\caption{(Color online) Energy levels as function of QW width with $k'_x = 0$ in Eq.~(\ref{eq:12}) (a) for different rotation angle $\alpha$ with respect to the anisotropy axes and taking the effective masses of monolayer BP, and (b) for a fixed angle $\alpha=\pi/4$ and assuming (solid curves) phosphorene and (dashed curves) arsenene parameters.} \label{fig:ExL}
\end{figure}

Let us consider the case of a QW with infinite potential walls ($V_0\rightarrow\infty$ in $V(y')$), for interfaces aligned along an arbitrary direction, i.e. non-zero $1/\mu_3$ (see Fig.~\ref{fig:anisotropic}). Without loss of generality, we will assume that the walls are parallel to the $x'$ direction, the assumed translational symmetry direction of the system, allowing us to write the wavefunction as $\Psi = \phi(y')e^{ik'_x x'}$. Using the Hamiltonian given by Eq.~(\ref{eq:3}) and the substitutions $\vec{p'}=\hbar\vec{k'}$ and $\vec{k'} \rightarrow -i\nabla'$, the resulting time-independent Schr\"odinger equation for the rotated QW becomes
\begin{equation}\label{eq:7}
-\frac{\hbar^2}{2\mu_2}\frac{d^2 \phi}{d y'^2} - i\frac{\hbar^2 k'_x}{\mu_3}\frac{d\phi}{dy'} + \frac{\hbar^2 k'^2_x}{2\mu_1}\phi = E\phi.
\end{equation}

We obtain a solution by assuming linear combinations of incident and reflected states as
\begin{align}\label{eq:8}
\Psi(x',y') = \left[A\exp (ik'^{+} _y y') + B \exp (i k'^{-} _y y')\right]e^{ik'_x x'},
\end{align}
with
\begin{subequations}\label{eq:9}
\begin{align}
& k'^{\pm}_y = \pm \theta_1 - \theta_2, \\
& \theta_1 = \sqrt{\frac{2\mu_2 E}{\hbar^2} - \left(\frac{\mu_2}{\mu_1} - \frac{\mu^2_2}{\mu^2_3}\right)k'^2_x}, \label{eq:9a}\\
& \theta_2 = \frac{\mu_2}{\mu_3}k'_x,\label{eq:9b}
\end{align}
\end{subequations}
where the plus (minus) sign refers to incident (reflected) waves. Now, one has to introduce the boundary conditions, i.e. the vanishing of $\phi$ at the interfaces, for $y'=0$ and $y'=L$ in Eq.~(\ref{eq:8}). That leads to the conditions $B=-A$ and $\sin(\theta_1 L) = 0$, resulting in the following quantization condition $\theta_1 = n\pi/L$ with $n\in \mathcal{Z}$. Therefore, the wavefunction and the energy levels are found as
\begin{subequations}
\begin{align}
&\Psi(x',y') = A \sin\left(\frac{n\pi}{L}y'\right)\exp\left[{i\left(x' - \frac{\mu_2}{\mu_3}y'\right)k'_x}\right], \label{eq:11} \\
& E = \frac{\hbar^2 n^2\pi^2}{2\mu_2 L^2} + \frac{\varrho\hbar^2 k'^2_x}{2\mu_1},\label{eq:12}
\end{align}
\end{subequations}
respectively, where $\varrho = 1-\frac{\mu_1 \mu_2}{\mu_3 ^2}$. It is seen that both the wavefunction, Eq.~(\ref{eq:11}), and the energy spectrum, Eq.~(\ref{eq:12}), show a striking dependence on the QW orientation $\alpha$ in relation to the anisotropy axes. 

\begin{figure}[!t]
\centerline{\includegraphics[width = \linewidth]{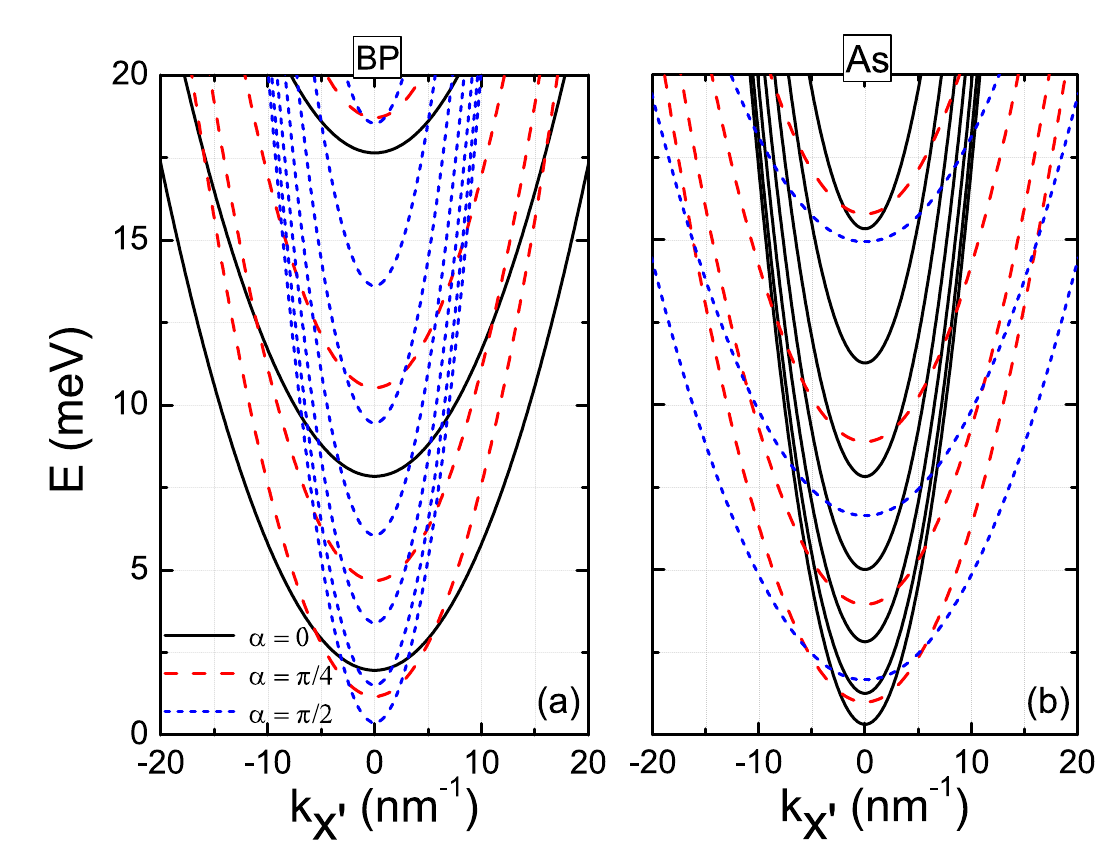}}
\caption{(Color online) Dispersion relation of QW with width $L=1$ nm for (a) phosphorene and (b) arsenene (As), and taking different rotation angle $\alpha$ with respect to the anisotropy axes.} \label{fig:ExK}
\end{figure}

\begin{figure}[!t]
\centerline{\includegraphics[width = 0.9\linewidth]{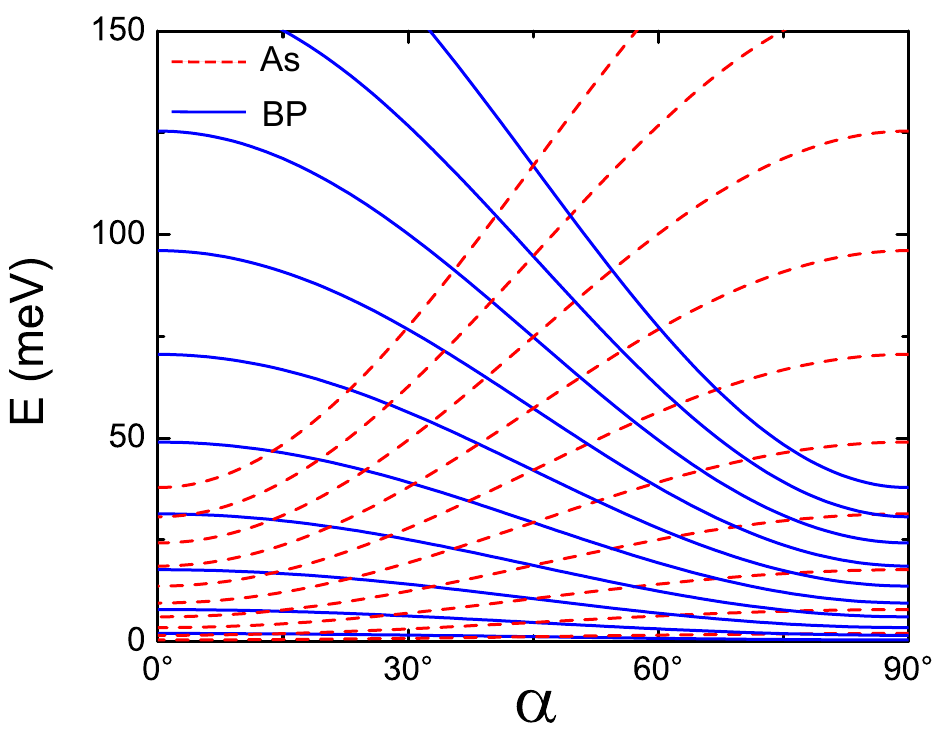}}
\caption{(Color online) Energy levels as a function of the rotation angle $\alpha$ with respect to the anisotropy axes (see Fig.~\ref{fig:anisotropic}) for (blue lines curves) phosphorene and (red dashed curves) arsenene QWs. It was taken $L=1$ nm and $k'_x = 0$.} \label{fig:Exalpha}
\end{figure}

Figure~\ref{fig:ExL} depicts the dependence of electronic energy levels of phosphorene and arsenene QWs with respect to the QW width $L$, by using Eq.~(\ref{eq:12}) with $k'_x=0$ and the effective masses of Table I. In panel (a) the energy levels for three different QW angles are shown for monolayer BP material, and in panel (b) we compare the electronic confined states of (solid curves) phosphorene and (dahsed curves) arsenene with the fixed angle $\alpha=\pi/4$. It is seen that the energy levels decrease quadratically with increasing QW width, something already expected when we make $k'_x=0$ in Eq.~(\ref{eq:12}), scaling as $\approx 1/L^2$ in a similar way as observed for confined states in $1$D squared quantum well and widely presented in quantum mechanic text books. A consequence of the change of QW alignment, as shown in Fig.~\ref{fig:ExL}(a), is a shift of the energy levels, together with a change of level spacing. By comparing the cases of a QW made of arsenene and phosphorene for a given value of rotation angle, shown in Fig.~\ref{fig:ExL}(b) it is seen that the behavior of the electronic levels of the two samples is similar. A difference that is evident in the dispersion relation in Figs.~\ref{fig:ExK}(a) and \ref{fig:ExK}(b), for QW with width $L = 1$ nm for BP and As, respectively, is the fact that the confined states in BP QWs present higher energy values ​​than those of arsenene. This is caused by the different effective masses of the materials (see Table I). Moreover, Fig.~\ref{fig:ExK} shows that, as $\alpha$ increases the energy levels are shifted to lower (upper) values for phosphorene (arsenene) and the spacing between them decreases (increases) too, which in turn increases (decreases) in the number of accessible electronic states. This result is emphasized in Fig.~\ref{fig:Exalpha}, which shows the energy levels as function of the alignment angle $\alpha$ for (blue solid curves) phosphorene and (dashed red curves) arsenene QW, maintaining the QW width $L = 1$ nm and $k'_x = 0$ in Eq.~(\ref{eq:12}). Note that these behaviors of the confined QW energy levels with respect to the rotation angle strongly resemble to those for 1D quantum well in Schr\"odinger equation with isotropic masses by varying the QW width instead of the alignment angle, i.e. the change of alignment QW angle with a fixed width $L$ for the anisotropic case works similarly as the isotropic case by varying the QW width. Phosphorene and arsenene energy levels exhibit opposite behaviors due to highest effective mass being along opposite directions in these materials (see Table~\ref{table}). These results suggest that a connection of QWs with different rotation angles acts similarly to constrictions in quantum point contact systems, due to the mismatch of the energy levels in the different sections of the QW junction. Such kind of QW junction system shall be explored latter in Sec.~\ref{sec.scattering}.      

\subsection{In the presence of magnetic field}\label{sec.B.ne.0}

Let us now study the effect of an external magnetic field perpendicular to the plane containing the QW, by considering the substitution $\vec{p'}\rightarrow \vec{p'}+e\vec{A}$ in Eq.~(\ref{eq:7}). A convenient choice of gauge is $\vec{A} = \left(-By', B\frac{\mu_2}{\mu_3}y', 0\right)$. In this case, one finds $\vec{\nabla} \cdot \vec{A} = (\mu_2/\mu_3) B$. Since we assume an uniform magnetic field, in this gauge the vector potential corresponds to an uniform rotation of the vector potential obtained from the Landau gauge by an angle of $\arctan(\mu_2/\mu_3)$. It is seen that for the isotropic case (i.e. $1/\mu_3 = 0$), as well as for $\alpha = 0$ and $\alpha = \pi/2$ one recovers the usual vector potential of the Landau gauge. Then, the Schr\"odinger equation for rotated QW in the presence of a perpendicular magnetic field can be written as
\begin{align}\label{eq:16}
\frac{-\hbar^2}{2\mu_2}\frac{d^2\Psi}{dy'^2} \hspace{-0.1cm}-\hspace{-0.1cm} i\frac{\hbar^2 k_x'}{\mu_3}\frac{d\Psi}{dy'} \hspace{-0.1cm}+\hspace{-0.1cm} \frac{\varrho}{2\mu_1}\hspace{-0.1cm}\left(\hspace{-0.05cm}eBy' \hspace{-0.1cm}+\hspace{-0.1cm} \hbar k_x'\hspace{-0.05cm}\right)\hspace{-0.1cm}^2 \hspace{-0.1cm}+\hspace{-0.1cm} (\hspace{-0.05cm}1 \hspace{-0.1cm}-\hspace{-0.1cm} \varrho \hspace{-0.05cm})\hspace{-0.05cm} \frac{\hbar^2 k'^2_x}{2\mu_1} \hspace{-0.05cm}=\hspace{-0.05cm} E\Psi.
\end{align}
Performing the coordinate transformation $y^* = y' + \frac{\hbar k_x'}{eB}$, defining the cyclotron frequency for the rotated anisotropic system as $w_c^2=\varrho\left(\frac{eB}{\mu_1}\right)^2$ and the new energies as $E'=E-(1-\varrho)\frac{\hbar^2 k_x'^2}{2\mu_1}$, one can rewrite Eq.~(\ref{eq:16}) as
\begin{eqnarray}\label{eq:17}
\frac{-\hbar^2}{2\mu_2}\frac{d^2\Psi}{dy^{*2}} - i\frac{\hbar^2 k_x'}{\mu_3}\frac{d\Psi}{dy^*} + \frac{\mu_1 w_c^2 y^{*2}}{2}\Psi = E'\Psi.
\end{eqnarray}

\begin{figure}[!t] 
\centerline{\includegraphics[width = 0.9\linewidth]{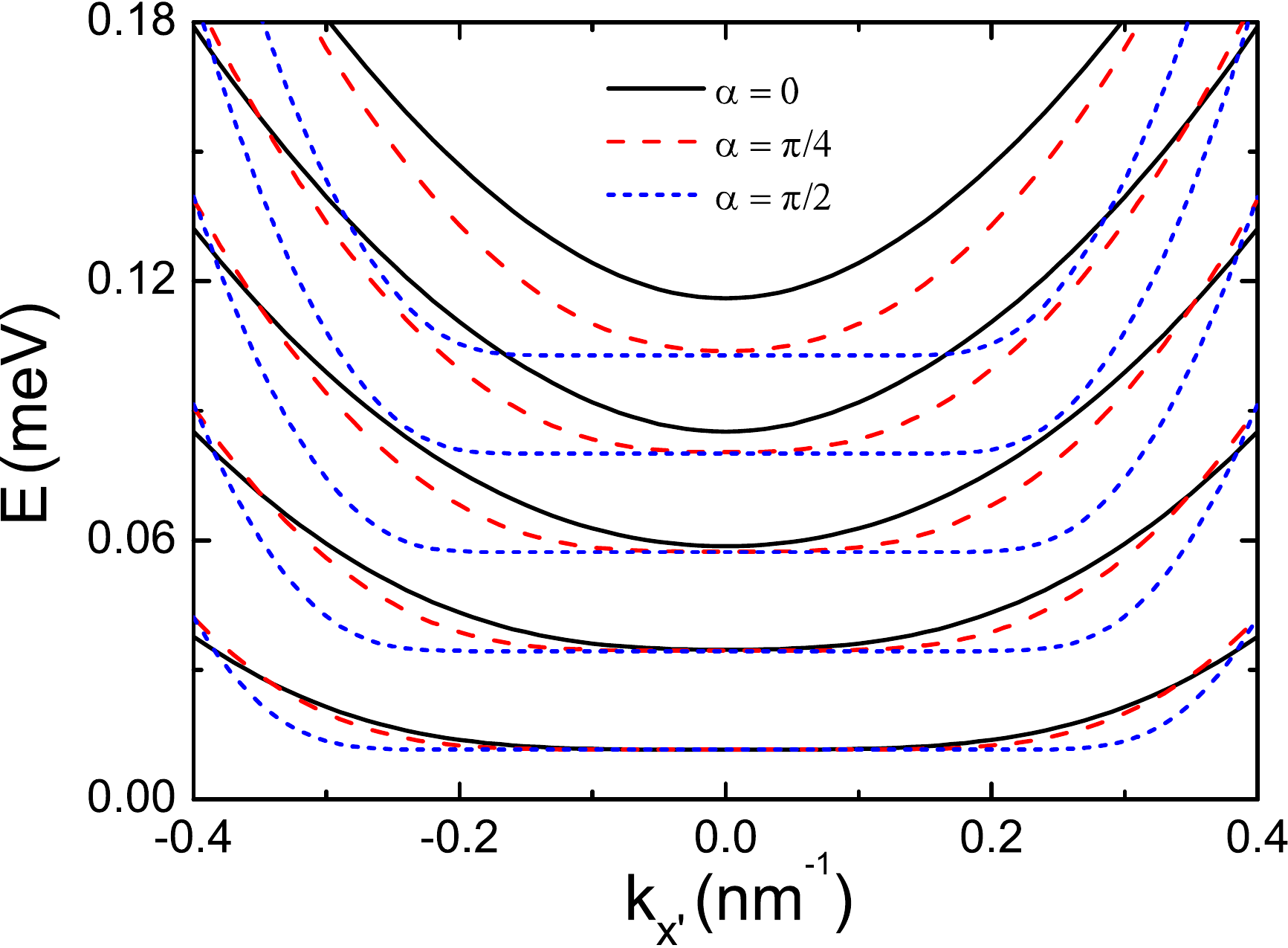}}
\caption{(Color online) Dispersion relation of phosphorene QW for different alignment angles $\alpha$ with respect to the anisotropy axes and fixed QW width $L = 100$ nm and magnetic field amplitude $B=5$ T. Black solid, red dashed, and blue short-dashed curves correspond to the spectrum for rotation angles $\alpha=0$, $\alpha=\pi/4$, and $\alpha=\pi/2$, respectively.} \label{fig:Energy1}
\end{figure}

\begin{figure}[!t]
\centerline{\includegraphics[width = 0.95\linewidth]{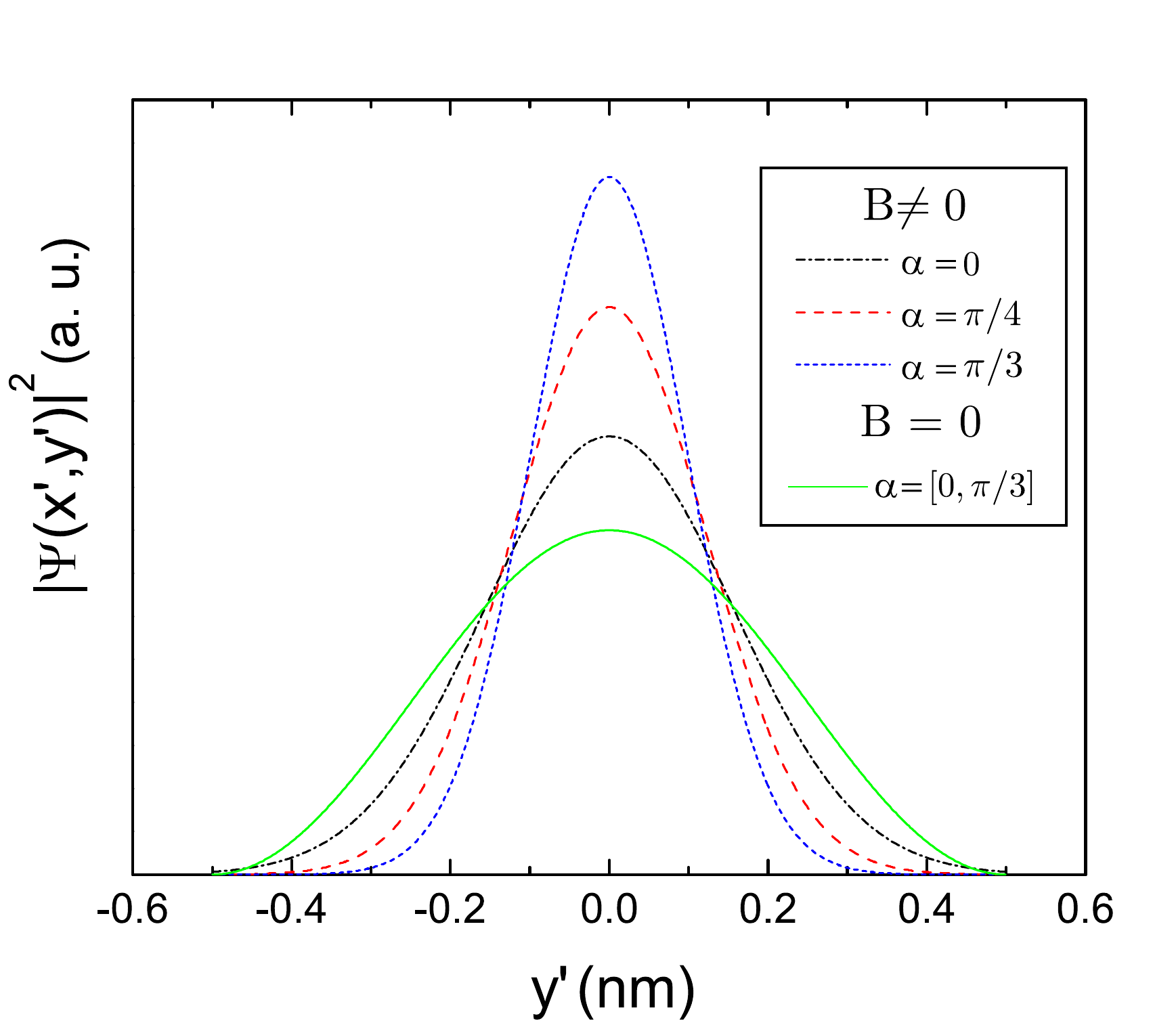}}
\caption{(Color online) Squared total wavefunction for anisotropic rotated QWs in the presence of an external magnetic field, $B=5$ T, with a fixed QW width $L = 100$ nm and wave vector $k'_x = 0$. Black dashed-dot, red dashed, and blue short-dashed curves correspond to the case for rotation angles $\alpha=0$, $\alpha=\pi/4$, and $\alpha=\pi/2$, respectively. For comparison, $|\Psi|^2$ for $B=0$ is shown by green solid curve.} 
\label{fig:bnonnull}
\end{figure}

By assuming the following ansatz $\Psi(x', y^*) = \exp\left(-i\mu_2 k_x' y^*/\mu_3\right)\phi(x', y^*)$ in order to eliminate the first derivative in Eq.~(\ref{eq:17}), it becomes
\begin{eqnarray}\label{eq:harmonic}
\frac{-\hbar^2}{2m^*}\frac{d^2\phi}{dy^{*2}} + \frac{m^*}{2} w_c^2 y^{*2}\phi = \sqrt{\frac{\mu_2}{\mu_1}}E\phi,
\end{eqnarray}
where $m^* = \sqrt{\mu_1\mu_2}$. Solving Eq.~(\ref{eq:harmonic}) numerically we obtain the energy levels for a QW in the presence of external magnetic field and different system parameters. Figure~\ref{fig:Energy1} shows the dispersion relation for different alignment angles (black solid curves) $\alpha=0$, (red dashed curves) $\alpha=\pi/4$, and (blue short-dashed curves) $\alpha=\pi/2$, and fixed QW width $L = 100$ nm and external magnetic field $B=5$ T. It is seen that similarly to an isotropic semiconductor structure with two boundaries, there is a momentum region around $k_x'=0$ where the energy levels are flat (i.e. $dE/dk_{x'}=0$). These states correspond to quantum Hall states, being more dispersive the higher the energy value, owing to the fact that the lower energy states are more strongly confined by the magnetic field. The presence of the edges gives rise to propagating states, resulting in the quantum Hall edge states. These states are related to the dispersive region of the energy spectrum in Fig.~\ref{fig:Energy1}, i.e. for momentum values away from the plateaus.[\onlinecite{edgestate1}-\onlinecite{quantumhall3}] In addition to the mentioned features, for the anisotropic QW case: (i) the quantum Hall edge states are significantly affected by the alignment of the QW, and (ii) as $\alpha$ increases, the energy states are found to be less dispersive, that is caused by the fact that the wavefunctions become more localized, as it will be discussed next in Fig.~\ref{fig:bnonnull}. Consequently, the group velocities of the quantum Hall edge states show a striking dependence on the edge alignment.  

In order to understand the effects of the rotation angle changes and the magnetic field on the electronic confined states, we show in Fig.~\ref{fig:bnonnull} the probability density of the ground state for different rotation angles with and without a magnetic field, taking the same system parameters as in Fig.~\ref{fig:Energy1}. Since Eq.~(\ref{eq:17}) is a quantum harmonic oscillator type equation, the ground state wavefunction of a rotated anisotropic QW in the presence of a magnetic field is given by
\begin{eqnarray}
\Psi(x',y*) \hspace{-0.075cm}=\hspace{-0.075cm} \left(\hspace{-0.075cm} \frac{m^* \omega_c}{\pi\hbar}\hspace{-0.075cm}\right)^{1/4} \hspace{-0.1cm}\exp\hspace{-0.075cm}\left(\hspace{-0.075cm}-\frac{m^*\omega_c y^{*2}}{2\hbar}\hspace{-0.05cm}-\hspace{-0.05cm}i\frac{\mu_2}{\mu_3}k'_x y^* \hspace{-0.075cm}\right). \ \ \ 
\end{eqnarray}
Similar to the case for zero magnetic field (see Eq.~(\ref{eq:11})), the wavefunction not only contains plane wave term but depends on the QW alignment in relation to anisotropy axes, which is contained into $y^*$ term. One can note that: (i) for a fixed rotation angle, the total wavefunction is more localized for $B\neq 0$ than $B=0$, as already expected, (ii) for $B\neq 0$, as $\alpha$ increases $|\Psi|^2$ becomes more localized, and (iii) for $B=0$, the QW rotations do not affect the wavefunction profile, as shown by the green solid curve in Fig.~\ref{fig:bnonnull} for $\alpha=0$ and $\alpha=\pi/3$.

\begin{figure}[!t]
\centerline{\includegraphics[width = \linewidth]{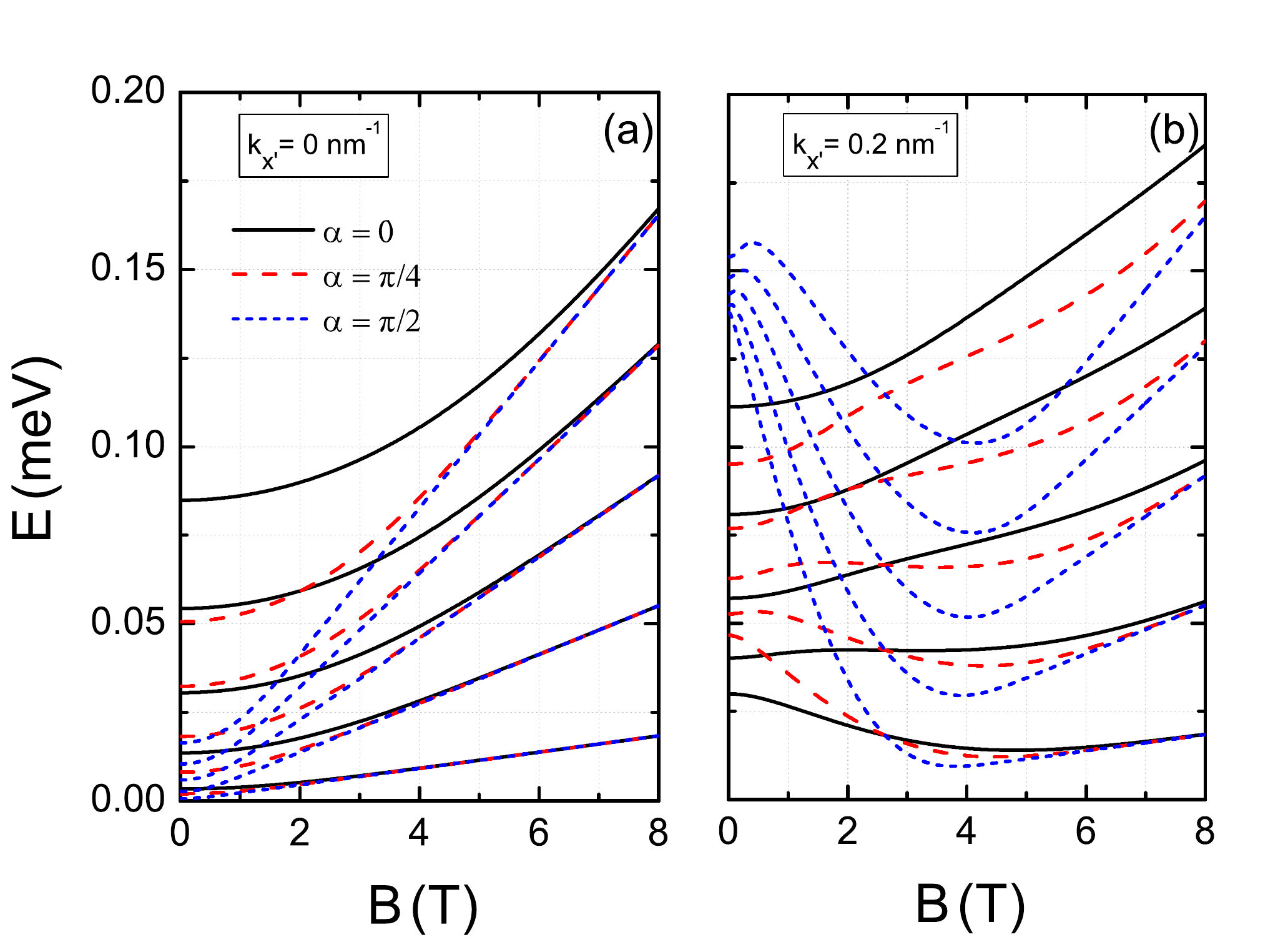}}
\caption{(Color online) Energy levels of a phosphorene QW with width $L = 100$ nm as function of the magnetic field for wave vector values (a) $k_x' = 0$, and (b) $k_x' = 0.2$ nm$^{-1}$ for different rotation angles $\alpha$ with respect to the anisotropy axes.} \label{fig:Energy2}
\end{figure}

A complementary way to see the magnetic field dependence of the confined states in anisotropic QWs is shown in Fig.~\ref{fig:Energy2}. The spectra for null and non-null wave vectors are present in panels (a) and (b), respectively, for three different rotation angles. Note that, as the magnetic field increases, the magnetic length becomes smaller than the system size, so that confinement effects are strongly reduced, and the magnetic levels in the phosphorene QW converge to the Landau levels of an infinite phosphorene sheet, given by: $E = \hbar\omega\left(n+1/2\right)$, with $n=0$, $1$, $2$, $\ldots$, and $\omega = eB/m_g = \omega_c\sqrt{\mu_1/\mu_2}$ being the cyclotron frequency calculated with the geometric mean of the masses $m_g=\sqrt{m_x m_y}$.~[\onlinecite{continuum0, continuum}] Moreover, one can realize that the energy levels spacing is strongly affected by the magnitude of the applied magnetic field, and with increasing magnetic field the confinement effects due to QW rotation discussed in Sec.~\ref{sec.B.eq.0} are less evident, such that regardless the wave vector amplitude (see, e.g. panels \ref{fig:Energy2}(a) and \ref{fig:Energy2}(b)) and QW rotation angle the energy levels converge to the Landau levels of an infinite system.  

\section{Wavepacket propagation and scattering in anisotropic QWs}\label{sec.scattering}

\begin{figure}[!t]
\centerline{\includegraphics[width = 0.75\linewidth]{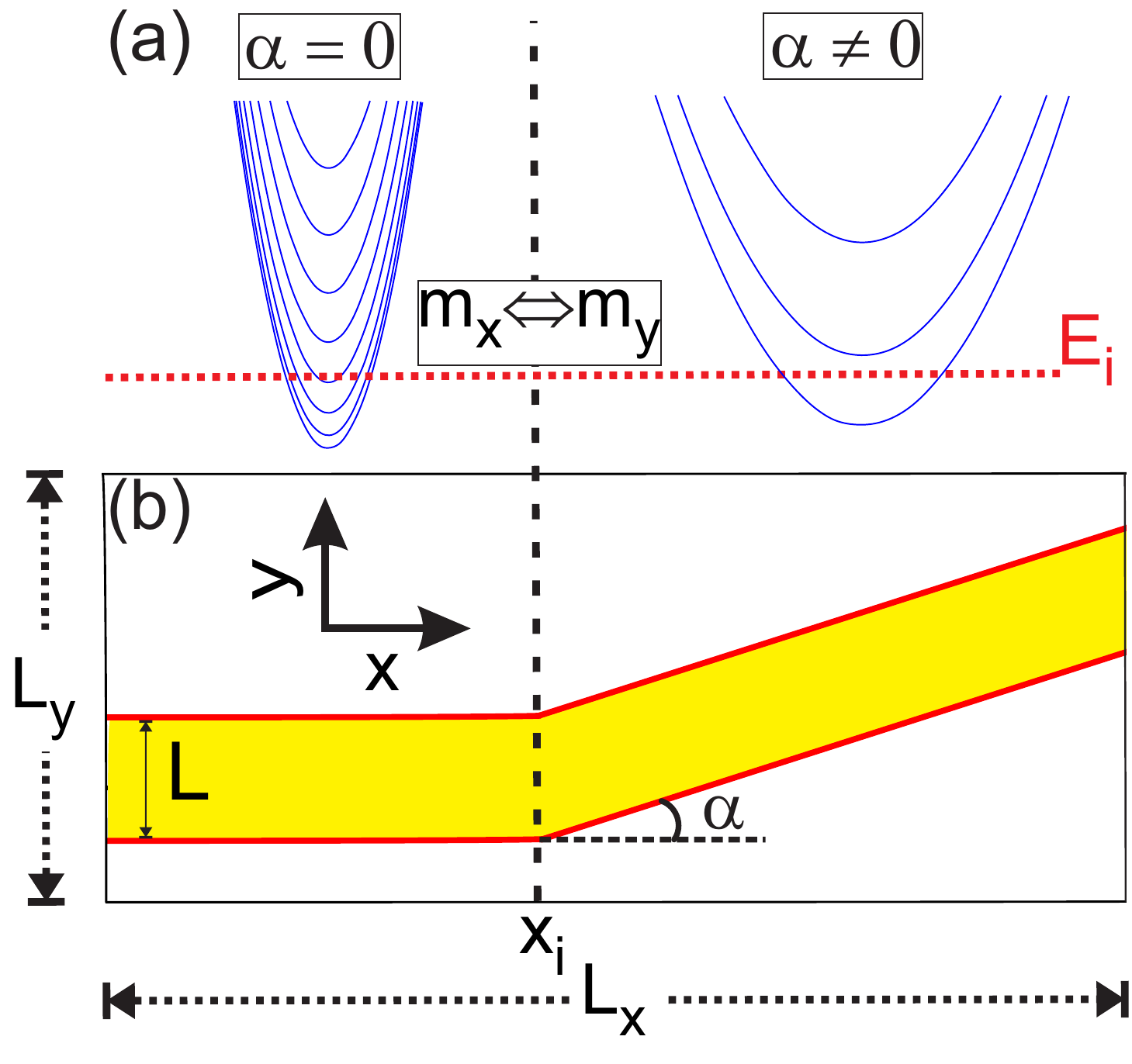}}
\caption{(Color online) Illustration of (a) the energy bands for each QW section of (b) the V-shaped anisotropic QW due to an abrupt change of orientation angle $\alpha$ along the QW length. For $x<x_i$ ($\geq x_i$), one has $\alpha=0$ ($\alpha\neq 0$). The two QW sections are made up of a phosphorene QW with width $L$. The energy bands for each QW section exhibit different energy levels spacing and minimum, and consequently leading to an energy mismatch in the junction. $E_i$ and $x_i$ indicate the initial wavepacket energy and the position of the QW corner. $L_x$ ($L_y$) is system length along the $x$ ($y$) direction.} \label{fig:quantum}
\end{figure}

As the previous results have shown, the electronic spectrum of anisotropic QWs is strongly dependent on the relative orientation of the QW in relation to the anisotropic axes. Therefore, it can be expected that a change of orientation angle ($\alpha$) along the length of the QW may give rise to an energy mismatch, as illustrated in Fig.~\ref{fig:quantum}(a), which can in turn lead to electron scattering. In order to investigate that, let us now calculate the transport properties of a QW in which an abrupt change of $\alpha$ is introduced, forming a elbow-like feature in an otherwise straight QW. For this purpose, let us now consider electrons in the $(x,y)$ plane moving from left to right in a region with a V-shaped QW formed by a straight section with $\alpha=0$ and a section with $\alpha\neq 0$ as illustrated in Fig.~\ref{fig:quantum}(b). The electrons are confined by a step like potential, i.e. $V(x,y) = 0$ inside the QW and $V(x,y) = V_0$ otherwise. Moreover, we assume that the electrons are always in the conduction band and that conduction-to-valence band transitions are negligible, which is a reasonable approximation when dealing with low-temperature systems and also once that the conduction-to-valence energy distance, i.e. the energy gap, is large for phosphorene systems [\onlinecite{continuum0, continuum, continuum1}]. It was considered QWs with width $L = 3$ nm and $L = 10$ nm, abrupt borders, and made out of phosphorene. For simplicity sake, throughout this section the effective masses along the $x$ and $y$ directions were exchanged as the ones referred in Table~\ref{table}.

The injected electrons are described by a combination of a Gaussian function with a plane wave along the $x$ direction and the ground state wavefunction of the QW in the $y$ direction $\phi_0(y)$. Then, at $t = 0$ the initial wavepacket is defined by
\begin{eqnarray}\label{eq:initialwave}
\Psi_0(x,y) = \exp\left[ik^i_xx - \frac{\left(x - x_0\right)^2}{2d^2}\right]\phi_0(y),
\end{eqnarray}
where $k^i_x = \sqrt{2m_x E_i/\hbar^2}$ is the wave vector corresponding to the packet kinetic energy $E_i$ (see dotted line in Fig.~\ref{fig:quantum}(a)), $d$ is the initial wavepacket width in the $x$ direction that is chosen as the same QW width $L$, and $x_0$ is the initial postion in the $x$ direction of the wave packet maximum, set up far from the corner of the bent QW, such as $x_0 = -32.5$~nm and $x_0 = -8.6$~nm for the QW width cases $L = 10$ nm and $L = 3$ nm, respectively. It is important to stress out that the ground state wavefunction $\phi_0(y)$ is given by Eq.~(\ref{eq:11}) which is closely related to the initial QW alignment angle, since it contains angle-dependent anisotropic effective masses terms. 

With the aim of solving the time-dependent Schr\"odinger equation to obtain the propagating wavepacket through the evolved time steps and thus to get the transport properties of the analyzed system, one applies the split-operator technique. For this, we follow the approach described in Refs.~[\onlinecite{SplitOperator2},\onlinecite{SplitOperator5}-\onlinecite{SplitOperator0},\onlinecite{SplitOperator1}-\onlinecite{SplitOperator13}]. This allows us to separate the exponential of the time evolution operator (that for the case in which the Hamiltonian does not explicitly depend on time, this operator can be written as $\hat{U}(t',t) = \exp\left[-\frac{i}{\hbar}H(t'-t)\right]$) into two parts: one of them involves only the potential operator $\hat{V}$, whereas the other contains only the kinetic operator $\hat{T}$, as well as, enabling to split also the kinetic terms for each direction. Therefore, the time evolved wavefunction is obtained by successively applying the operation $\hat{U}$ such as
\begin{align}
\Psi(\hspace{-0.05cm}\vec{r},t+\Delta t\hspace{-0.05cm}) \hspace{-0.1cm}=\hspace{-0.1cm} e^{\hspace{-0.075cm} -i\hat{V}\hspace{-0.075cm}\Delta t/2\hbar}e^{-i\hat{T}_x\Delta t/\hbar}e^{\hspace{-0.075cm}-i\hat{T}_y\hspace{-0.075cm}\Delta t/\hbar}e^{\hspace{-0.075cm}-i\hat{V}\Delta t/2\hbar}\Psi(\hspace{-0.05cm}\vec{r},t\hspace{-0.05cm}),
\end{align}
where $\hat{T}_{x(y)}$ is the kinetic-energy operator for $x(y)$ direction and we neglect terms of order $\mathcal{O}\left(\Delta t^3\right)$ and higher, being such error a consequence of the noncommutativity of kinetic and potential terms. This error can be minimized as smaller the time step. We assume a small time step of $\Delta = 0.7$ fs. Here, we opted for the split-operator technique, because it allows us to track the position and velocity of the center of mass trajectories, see reflection patterns and scattering on the edges, and obtain the transmission and reflection coefficients (which will be important to the analysis in this section).

\begin{figure}[!t]
\centerline{\includegraphics[width = 0.9\linewidth]{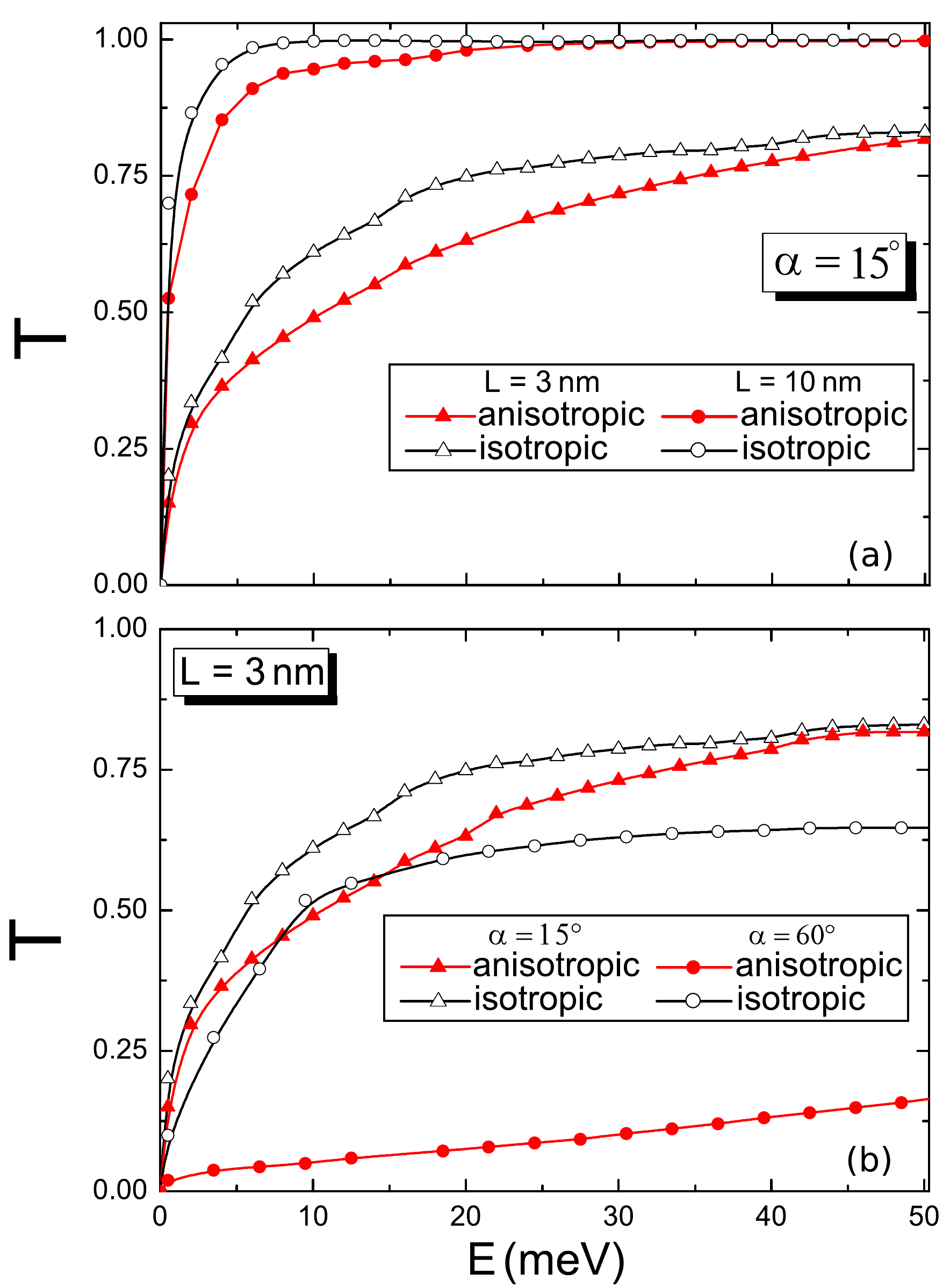}}
\caption{(Color online) Transmission probabilities as a function of the initial wavepacket energy by assuming the elbow-like QW with: (a) a fixed rotation angle $\alpha = 15^\circ$ and QW widths $L = 3$ nm (triangles) and $L = 10$ nm (circles); and (b) a fixed QW width $L = 3$ nm and rotation angles $\alpha = 15^\circ$ (triangles) and $\alpha = 60^\circ$ (circles). Red (filled) and black (open) curves (symbols) correspond to the anisotropic and isotropic QW cases.} \label{fig:transmission}
\end{figure}

To numerically solve this problem, we discretized the $(x, y)$ plane with a square grid, assuming $\Delta x = \Delta y = 0.4$ nm and $\Delta x = \Delta y = 0.12$ nm for the cases where $L = 3$ nm and $L = 10$ nm, respectively, and used the finite difference scheme to solve the derivatives in the kinetic energy terms of the Hamiltonian. In addition, as suggested in Ref.~[\onlinecite{manolopoulos}] and successfully used in Refs.~[\onlinecite{SplitOperator3, SplitOperator4, SplitOperator6, SplitOperator13}] we added an absorbing (imaginary) potential on the boundaries of our computational box in order to avoid spurious reflections and backscattering when the wavepacket reaches the limits of our system.

For each investigated system configuration, we run the simulation and calculate: (i) the transmission probability $T(t)$ for each time step by integrating the square modulus of the normalized wavepacket in the region after the elbow-like QW corner, i.e. for $x>x_i$, given by 
\begin{align}\label{eq.T}
T(t) = \int_{-L_y/2}^{L_y/2}dy\int_{x_i}^{L_x/2-|x_i|}dx\left|\Psi(x,y,t)\right|^2,
\end{align}
(ii) the total average position, i.e., the trajectory of the wavepacket center of mass, that is calculated for each time step by computing
\begin{subequations}
\begin{align}
\langle x(t)\rangle &= \int_{-L_y/2}^{L_y/2}dy\int_{-L_x/2}^{L_x/2}dx\left|\Psi(x,y,t)\right|^2 x,\label{eq.X}\\
\langle y(t)\rangle &= \int_{-L_x/2}^{L_x/2}dx\int_{-L_y/2}^{L_y/2}dy\left|\Psi(x,y,t)\right|^2 y,\label{eq.Y}
\end{align}
\end{subequations}
and (iii) the average velocity, by 
\begin{subequations}
\begin{align}
\langle v_x(t)\rangle &= \frac{d\langle x(t)\rangle}{dt},\label{eq.VX}\\
\langle v_y(t)\rangle &= \frac{d\langle y(t)\rangle}{dt}\label{eq.VY},
\end{align}
\end{subequations}
where the limits of the computational box are defined by $x\in [-L_x/2,L_x/2]$ and $y\in [-L_y/2,L_y/2]$. The reflection probability $R$ is obtained by similar integration as Eq.~(\ref{eq.T}) but for the region before the QW corner ($x<x_i$). For larger $t$, the value of the transmission (reflection) probability integral increases (decreases) with time until it converges to a number. This number is then considered to be the transmission (reflection) probability of such a system configuration.

\begin{figure*}[!t]
\centering
\includegraphics[width = \linewidth]{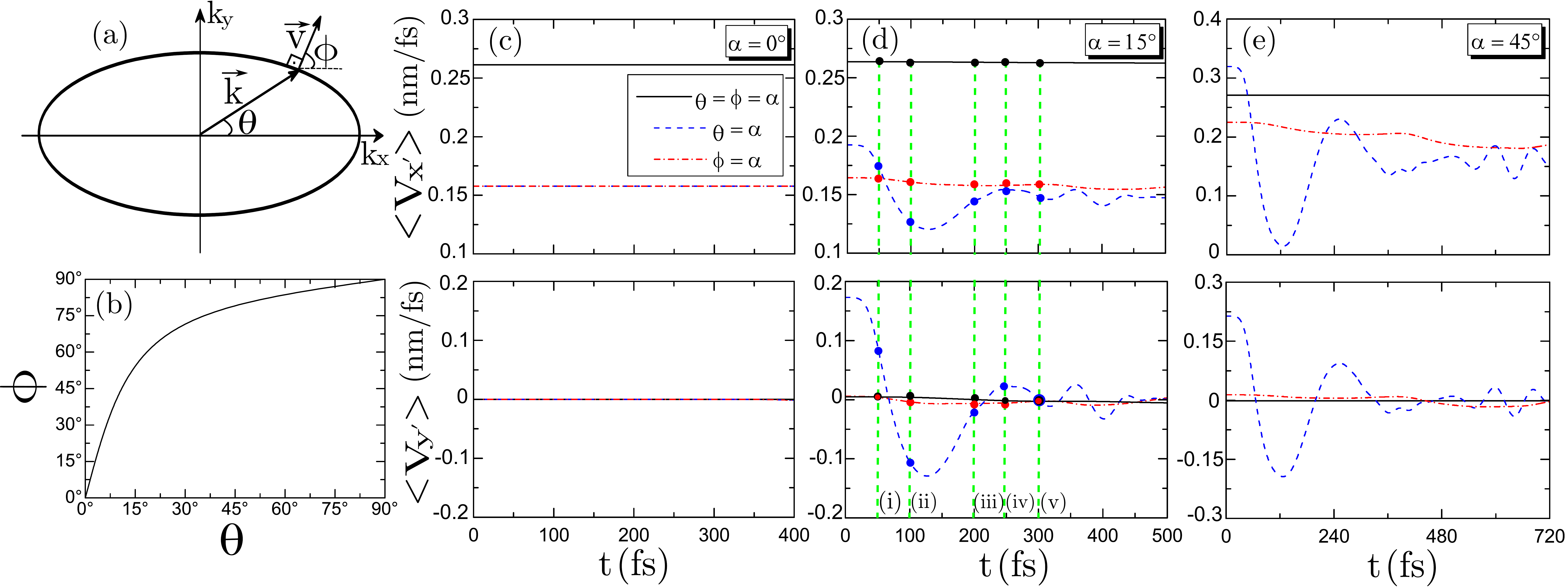}
\caption{(Color online) (a) Isoenergy curve in momentum space of the electronic band for a non-rotated anisotropic system, that corresponds to $\alpha=0$. $\theta$ and $\phi$ are the minimum angles with respect to $k_x$ axis that are associated with the orientation of the wave vector $\vec{k}$ and group velocity vector $\vec{v}$, respectively. (b) Values of angle $\phi$ as a function of angle $\theta$ given by the following equation: $\tan\phi=\left(\tan\theta/\mu_2-1/\mu_3\right)/\left(\tan\theta/\mu_3+1/\mu_1\right)$. (c)-(e) Average velocities for the (top panels) $x'$ and (bottom panels) $y'$ directions by considering (black solid curve that corresponds to $\theta=\phi=\alpha$) the isotropic case, the anisotropic case with the QW parallel to (blue dashed curve that corresponds to $\theta=\alpha\neq\phi$) the wave vector, and (red dashed-dot curve that corresponds to $\phi=\alpha\neq\theta$) to the group velocity vector. The rotation angle was assumed as (c) $\alpha=0^\circ$, (d) $\alpha=15^\circ$, and (e) $\alpha=45^\circ$.}
\centering
\label{velocity}
\end{figure*}

\begin{figure*}[!btph]
\centering
\includegraphics[width = \linewidth]{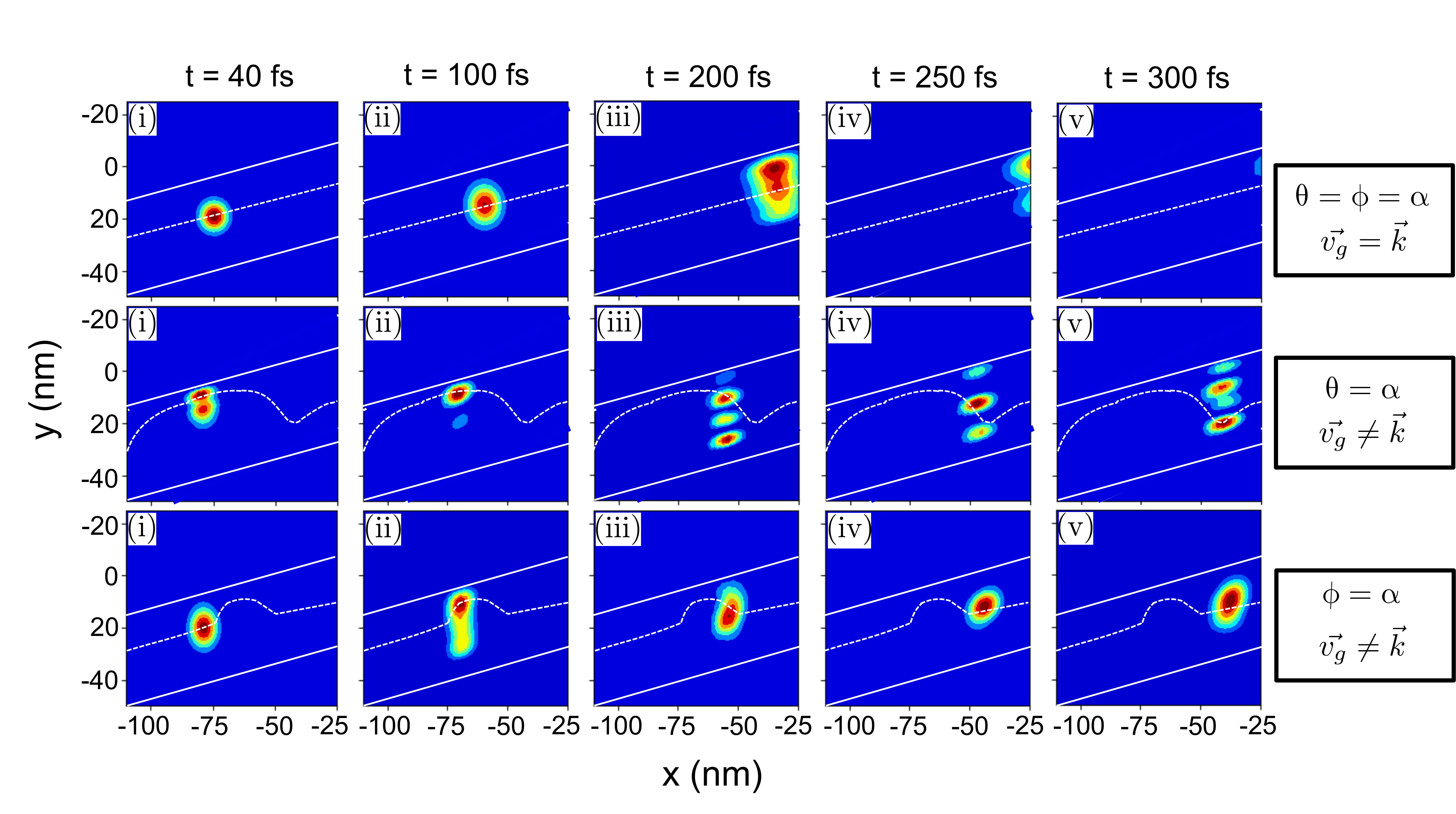}
\caption{(Color online) Snapshots of the total evolved wavefunction through the QW rotated by $\alpha = 15^\circ$ at the time steps (i) $t = 40$~fs, (ii) $t = 100$~fs, (iii) $t = 200$~fs, (iv) $t = 250$~fs, and (v) $t = 300$~fs as labeled by roman letters in Fig.~\ref{velocity}(d) and considering (upper panels) $\theta = \phi = \alpha$, (middle panels) $\theta = \alpha$, and (bottom panels) $\phi = \alpha$.}
\centering
\label{snapshots}
\end{figure*}

Transmission probabilities for the bent QW computed by using the split-operator technique are presented in Fig.~\ref{fig:transmission} as function of the initial wavepacket energy. In Fig.~\ref{fig:transmission}(a) the transmission was obtained for a QW rotated by a fixed angle $\alpha=15^\circ$ and QW width $L = 3$~nm (triangles) and $L = 10$~nm (circles) both in isotropic (open symbols) and anisotropic cases (filled symbols). In Fig.~\ref{fig:transmission}(b), it was fixed the QW width $L = 3$~nm and analyzed two different rotation angles: $\alpha = 15^\circ$ (triangles) and $\alpha = 60^\circ$ (circles). From Fig.~\ref{fig:transmission}(a), one can notice that: (i) since the energy levels become closer for wider QWs, the wavepacket has a larger transmission probability for wider channels in both isotropic and anisotropic systems, as well as, it also explains the rapid convergence of the transmission to $1$ for wider QWs as a consequence of the larger number of accessible electronic states; (ii) the quantitative difference between anisotropic and isotropic curves for each fixed QW width case is due to the difference on their subbands energy values. Note that the energy bands in both straight and rotated sections of the V-shaped QW are (non-)identical for (an)isotropic case, and thus as a consequence of this energy mismatch caused by the QW bending one has a greater reflection probability for anisotropic case. In both isotropic and anisotropic cases, due to the channel geometry the wavepacket is more reflected when reaches the bend that connects the two leads represented in Fig.~\ref{fig:quantum}(b), and as the right-arm of QW is rotated the transmission decreases, as can be seen in Fig.~\ref{fig:transmission}(b). The introduction of a bend in the QW can reduce the transmission even in the isotropic case, due to the fact that it breaks the translational symmetry of the system. For instance, compare black triangular and circular symbols in Fig.~\ref{fig:transmission}(b), in which the transmission for $\alpha=15^\circ$ is larger than the case for $\alpha=60^\circ$ for any initial wavepacket energy. Although the energy bands for the isotropic case are identical for any rotation angles, the QW geometry has an important role on the total transmission probability. Thus, in order to separate this purely geometric effect from the effect of the anisotropy, all the results in Fig.~\ref{fig:transmission} show a comparison between the transmission for isotropic and anisotropic cases for different values of $L$ (Fig.~\ref{fig:transmission}a) and $\alpha$ (Fig.~\ref{fig:transmission}b).

Another aspect of the transport in anisotropic QWs that was investigated was the effect of the interaction between the electrons and the QW edges as function of $\alpha$. A semiclassical analysis suggests that the non-collinearity of the group velocities and the momentum vectors (see Eq.~(\ref{eq:5.1})) may give rise to a group velocity oscillation. In order to investigate that, we have analyzed the wavepacket dynamics simulating electrons that propagate through a straight QW as the one represented in Fig.~\ref{fig:anisotropic}. For this, we considered an initial circularly symmetric Gaussian wavepacket centered in $\vec{r} = (x_0,y_0)$ and multiplied by a pulse with initial wave vector $\vec{k}_0$, given by
\begin{eqnarray}\label{eq:gaussian}
\Psi(x,y) = \exp\left[-\frac{(x-x_0)^2}{2d^2} - \frac{(y - y_0)^2}{2d^2} + i\vec{k_0}\cdot\vec{r}\right].
\end{eqnarray}
In this analysis, it was assumed that the QW width is much larger than the wavepacket width, being taken $d=5$~nm and $L=30$~nm, and that the wavepacket is injected from left to right into the channel with initial position $(x_0,y_0) = (-32.5,-8.6)$~nm and initial energy $E = 200$~meV. 

Figure~\ref{velocity} shows the wavepacket average velocities as function of time for both $x$ and $y$ directions that were obtained by computing the first order derivative of the average positions Eqs.~(\ref{eq.X}) and (\ref{eq.Y}) at each time step, being given by Eqs.~(\ref{eq.VX}) and (\ref{eq.VY}), respectively. The wavepacket evolution through the straight QW with different values of rotations angle $\alpha$ was analyzed for both isotropic and anisotropic cases, taking into account the non-collinearity of the wave vector $\vec{k}$ and group velocity vector $\vec{v}$. The wave vector and group velocity are here associated with the angles $\theta$ and $\phi$, respectively, as illustrated in Fig.~\ref{velocity}(a), being $\vec{v}$ always perpendicular to the isoenergy in momentum space. It is easy to see from Eq.~(\ref{eq:5.1}) that for isotropic case ($1/\mu_3=0$), one has $\vec{v}'\parallel\vec{k}'$ and the isoenergies are circular. However, as mentioned in Sec.~\ref{sec.classic}, for anisotropic semiconductors whose isoenergies are ellipses, this is not the case. Figure~\ref{velocity}(b) shows the relation between the angles $\theta$ and $\phi$ that differs for almost every angle, except for $\theta = 0^\circ$ and $\theta = 90^\circ$ in which the wave vector and group velocity are aligned, similarly to the isotropic case. Figures~\ref{velocity}(c) to \ref{velocity}(e) depict the average velocities (top panels) $v'_x$ and (bottom panels) $v'_y$ for the following rotation angles: [Fig.~\ref{velocity}(c)] $\alpha=0^\circ$, [Fig.~\ref{velocity}(d)] $\alpha=15^\circ$, and [Fig.~\ref{velocity}(e)] $\alpha=45^\circ$. The black solid, blue dashed, and red dashed-dot curves correspond to $\theta=\phi=\alpha$, i.e. the isotropic case, to $\theta=\alpha\neq\phi$, i.e. the anisotropic case with the QW parallel to the wave vector, and to $\phi=\alpha\neq\theta$, i.e. the anisotropic case with the QW parallel to the group velocity vector, respectively. According to Figs.~\ref{velocity}(c)-\ref{velocity}(e), one can realize that: (i) the average velocities for both $x$ and $y$ directions remain unchanged for isotropic case ($\theta=\phi=\alpha$, black solid curves), irrespective to the QW rotation angle, as well as for the anisotropic case in which the wave vector and the group velocity are collinear as shown by the blue dashed and red dashed-dot curves in Fig.~\ref{velocity}(c). Qualitative similar results can be obtained for $\alpha=90^\circ$, instead of $\alpha=0^\circ$; (ii) for $\theta\neq\phi$ and $\alpha\neq 0^\circ,90^\circ$ that corresponds to non-collinear cases between the wave vector and the group velocity, the average velocities oscillate, as expected by the semiclassical picture due to the non-specular reflections on the edges in an anisotropic media. This can be seen by the blue dashed and red dashed-dot curves in Figs.~\ref{velocity}(d) and \ref{velocity}(e); (iii) the oscillations are more evident as $|\theta-\phi|$ increases, exhibiting an increasing oscillation amplitude the greater the non-collinearity between the vectors $\vec{k}$ and $\vec{v}$. This can be seen by comparing the oscillation amplitudes of the blue dashed curves in Figs.~\ref{velocity}(d) and \ref{velocity}(e), and also for a fixed rotation angle by comparing the $\theta=\alpha$ and $\phi=\alpha$ cases. Note from Fig.~\ref{velocity}(b) that for $\theta = \alpha = 15^\circ$ one has $\phi\approx 55^\circ$, whereas for $\phi = \alpha = 15^\circ$ one implies $\theta \approx 4^\circ$, and consequently the difference $|\theta-\phi|$ is larger for former case with $\theta=\alpha$ (blue dashed curves) that indeed exhibits the large oscillation amplitude for the presented cases.

In order to clarify how the non-collinearity between the group velocity and wave vector in anisotropic case affects the wavepacket evolution, it is displayed in Fig.~\ref{snapshots} snapshots of the time evolution of the probability density propagating through the QW rotated by the angle $\alpha = 15^\circ$ at times (i) $t = 40$~fs, (ii) $t = 100$~fs, (iii) $t = 200$~fs, (iv) $t = 250$~fs, and (v) $t = 300$~fs as labeled in Fig.~\ref{velocity}(d), and considering the isotropic case (upper panels, $\theta = \phi = \alpha$) and the anisotropic case with the QW orientation parallel to the wave vector (middle panels, $\theta = \alpha\neq\phi$) and to the group velocity (bottom panels, $\phi=\alpha\neq\theta$). By analyzing the snapshots, it is clear that for the isotropic case (upper panels) when the wavepacket evolves it disperses but keeping the average position (white dashed lines) and consequently the average velocity unchanged, as observed in Fig.~\ref{velocity}(d). Since the propagation direction and the wave vector are collinear for this case, after the reflections at the potential edges the direction of the group velocity vector remains the same over time. However, for the anisotropic case in which the wave vector and the group velocity are non-collinear, when the wavepacket reaches the QW edges it undergoes non-specular reflections [\onlinecite{nonspecular}]. As a consequence, for the case where $\theta=\alpha$, this interaction with the edges results a subpackage splitting with different propagation directions that leads to an average velocity oscillation with large amplitudes that are damped over time, as shown by the blue dashed curves in Fig.~\ref{velocity}(d). On the other hand, for the anisotropic case where $\phi=\alpha$, no subpackage splitting is observed and the average velocity oscillation amplitude is less pronounced, as shown by red dashed-dot curves in Fig.~\ref{velocity}(d). This is linked to the fact that in this case the group velocity is aligned with the QW orientation and then the total wavepacket evolves in parallel to QW boundaries exhibiting a straight trajectory and dispersing over time similarly to the isotropic snapshots case, but here owing to the non-specular reflections its interaction with the QW edges implies a slightly different average position and barely affecting the total propagation velocity.

\section{Conclusions}\label{sec.conclusions}

In summary, we developed an analytical model for classical anisotropic systems using the effective mass model and applied this formalism to obtain the electronic properties of QWs made up of arsenene and phosphorene and with the length direction rotated in relation to its anisotropy axes. The energy levels in the presence and absence of an external magnetic field perpendicular to the QW plane were analyzed for different system parameters. In the absence of a magnetic field, we found an analytical expression for the QW energy levels that contains a term analog to the ones for isotropic quantum wells with a $1/L^2$ dependence, and another term that carries the system anisotropy. Our results showed that the spacing of the energy levels for both samples is strongly affected by the alignment angle between the QW and the crystallographic directions, such that as the angle increases, the spacing between the energy levels is lowered (raised) for phosphorene (arsenene), as well as, observing a shifted to lower (upper) energy values. For the non-null magnetic field case, the electronic wavefunctions obey a harmonic oscillator type equation but for a modified mass and modified cyclotron frequency that depends on the alignment angle between the QW and its anisotropy axes. Numerical calculations showed that the energy spectrum is significantly affected by the confining potential edges and that the quantum Hall edge states are less pronounced the greater the rotation angle. With respect to the wavefunction localization, for large QW rotation angles, the wavefunction becomes more confined, whereas in the absence of a magnetic field it remains unchanged under rotations.

Since the electronic energy levels of anisotropic QWs are strongly affected by rotation, we studied their transport properties by using the split-operator technique and compared the isotropic and anisotropic results for the transmission probability, average position, average group velocity, and snapshots of the time evolved wavepacket. By considering a circularly symmetric Gaussian wavepacket propagating inside of a wide anisotropic QW rotated by $\alpha$ with respect to the anisotropic axes, one observed oscillations in the average velocity for the case when the initial wave vector and the group velocity vector are not collinear, and the oscillation amplitude is more pronounced the greater the non-collinearity between them, i.e. the greater the $\theta-\phi$ value. The snapshots at different time steps demonstrated that for the anisotropic QWs the interaction between the wavepacket and the QW edges gives rise to subwavepackets with different momentum orientations, whereas for isotropic QWs the wavepacket disperses over time without splitting and its interaction with the QWs edges does not change the orientation of the average group velocity. In the case of a bent QW, as a consequence of the energy mismatching in different sections of the QW and the anisotropy of the system, one expects that electrons traveling through the bend can be scattered. The results showed that the transmission probabilities are greater the lower the rotation angle of the right-arm and the wider the QW, regardless of the anisotropic character of the system, and the nature of the quantitative difference of the transmission probabilities between the isotropic and anisotropic QWs is linked to the difference on their subband values. The differences in propagation for different orientations of the QW may be experimentally measured by attaching perpendicular leads to the system, one expecting different Hall conductances between isotropic and anisotropic cases, as well for collinear and non-collinear situations between the group velocity and momentum vectors. This direction-dependent Hall conductance will be investigated in a future project. Finally, we hope that our electronic and transport results will prove useful for designing anisotropic semiconductor based quantum confinement devices.

\section*{ACKNOWLEDGMENTS}
This work was financially supported by the Brazilian Council for Research (CNPq), under the PRONEX/FUNCAP and CAPES foundation.


\begin{references}

\bibitem{graphene1} K. S. Novoselov, A. K. Geim, S. V. Morozov, D. Jiang, Y. Zhang, S. V. Dubonos, I. V. Grigorieva, and A. A. Firsov, Electric field effect in atomically thin carbon films, Science \textbf{306}, 666 (2004).

\bibitem{graphene2} K. S. Novoselov, A. K. Geim, S. V. Morozov, D. Jiang, M. I. Katsnelson, I. V. Grigorieva, S. V. Dubonos, and A. A. Firsov, Two-dimensional gas of massless Dirac fermions in graphene, Nature (London) \textbf{438}, 197 (2005).

\bibitem{graphene3} A. H. C. Neto, F. Guinea, N. M. R. Peres, K. S. Novoselov, and A. K. Geim, The electronic properties of graphene, Rev. Mod. Phys. \textbf{81}, 109 (2009).

\bibitem{graphene4} M. I. Katsnelson, Graphene: Carbon in Two Dimensions (Cambridge University Press, Cambridge, 2012).

\bibitem{phosphorene6} P. Avouris, T. F. Heinz, and T. Low, 2D Materials (Cambridge University Press, 2017).

\bibitem{layered1} K. Novoselov, D. Jiang, F. Schedin, T. J. Booth, V. V. Khotkevich, S. V. Morozov, and A. K. Geim, Two-dimensional atomic crystals, Proc. Natl. Acad. Sci. U.S.A. \textbf{102}(30), 10451 (2005).

\bibitem{layered2} Q. H. Wang, K. Kalantar-Zadeh, A. Kis, J. N. Coleman, and M. S. Strano, Electronics and optoelectronics of two-dimensional transition metal dichalcogenides, Nat. Nanotechnol. \textbf{7}, 699 (2012).

\bibitem{layered3} M. Xu, T. Liang, M. Shi, and H. Chen, Graphene-like twodimensional materials, Chem. Rev. \textbf{113}, 3766 (2013).

\bibitem{layered4} E. Bianco, S. Butler, S. Jiang, O. D. Restrepo, W. Windl, and J. E. Goldberger, Stability and exfoliation of germanane: A germanium graphane analogue, ACS Nano \textbf{7}(5), 4414 (2013).

\bibitem{layered5} P. Vogt, P. D. Padova, C. Quaresima, J. Avila, E. Frantzeskakis, M. C. Asensio, A. Resta, B. Ealet, and G. L. Lay, Silicene: Compelling experimental evidence for graphene like two-dimensional silicon, Phys. Rev. Lett. \textbf{108}(15), 155501 (2012). 

\bibitem{layered6} M. Chhowalla, H. S. Shin, G. Eda, L.-J. Li, K. P. Loh, and H. Zhang, The chemistry of two-dimensional layered transition metal dichalcogenide nanosheets, Nat. Chem. \textbf{5}, 263 (2013).

\bibitem{continuum0} D. J. P. de Sousa, L. V. de Castro, D. R. da Costa, and J. M. Pereira, Boundary conditions for phosphorene nanoribbons in the continuum approach, Phys. Rev. B \textbf{94}, 235415 (2016).

\bibitem{SplitOperator2} M. H. Degani and M. Z. Maialle, Numerical Calculations of the Quantum States in Semiconductor Nanostructures, J. Comput. Theor. Nanosci. \textbf{7}, 454 (2010).

\bibitem{SplitOperator5} M. D. Petrovi\'c, F. M. Peeters, A. Chaves, and G. A. Farias, Conductance maps of quantum rings due to a local potential perturbation, J. Phys.: Condens. Matter \textbf{25}, 495301 (2013).

\bibitem{confine} M. Zarenia, A. Chaves, G. A. Farias, and F. M. Peeters, Energy levels of triangular and hexagonal graphene quantum dots: A comparative study between the tight-binding and Dirac equation approach, Phys. Rev. B \textit{84}, 245403 (2011).

\bibitem{confine2} Z. Wu, Z. Z. Zang, K. Chang, and F. M. Peeters, Quantum tunneling through graphene nanorings, Nanotechnology \textit{21}, 185201 (2010).

\bibitem{confine3} M. Grujić, M. Zarenia, A. Chaves, M. Tadić, G. A. Farias, and F. M. Peeters, Electronic and optical properties of a circular graphene quantum dot in a magnetic field: Influence of the boundary conditions, Phys. Rev. B \textit{84}, 205441 (2011).

\bibitem{confine4} P. Hewageegana and V. Apalkov, Electron localization in graphene quantum dots. Phys. Rev. B \textit{77}, 245426 (2008).

\bibitem{confine5} C. A. Downing, D. A. Stone, and M. E. Portnoi, Zero-energy states in graphene quantum dots and rings, Phys. Rev. B \textit{84}, 155437 (2011).

\bibitem{confine6} J. Schelter, B. Trauzettel, and P. Recher, How to distinguish between specular and retroconfigurations for Andreev reflection in graphene rings, Phys. Rev. Lett. \textit{108}, 106603 (2012).

\bibitem{confine7} S. Zhang, H. Chen, E. Zhang, and D. Liu, The Aharonov-Anandan current induced by a time-dependent magnetic flux in graphene rings, EPL \textit{103}, 58005 (2013).

\bibitem{continuum} J. M. Pereira and M. I. Katsnelson, Landau levels of single-layer and bilayer phosphorene, Phys. Rev. B \textbf{92}, 075437 (2015).

\bibitem{continuum1} D. J. P. de Sousa, L. V. de Castro, D. R. da Costa, J. M. Pereira, and T. Low, Multilayered black phosphorus: From a tight-binding to a continuum description, Phys. Rev. B \textbf{96}, 155427 (2017).

\bibitem{phosphorene1} L. Li, Y. Yu, G. J. Ye, Q. Ge, X. Ou, H. Wu, D. Feng, X. H.
Chen, and Y. Zhang, Black phosphorus field-effect transistors, Nat. Nanotechnol. \textbf{9}, 372 (2014).

\bibitem{phosphorene2} H. Liu, A. T. Neal, Z. Zhu, Z. Luo, X. Xu, D. Tománek, and P. D. Ye, Phosphorene: An unexplored 2D semiconductor with a high hole mobility, ACS Nano \textbf{8}, 4033 (2014).

\bibitem{phosphorene3} F. Xia, H. Wang, and Y. Jia, Rediscovering black phosphorus as an anisotropic layered material for optoelectronics and electronics, Nat. Commun. \textbf{5}, 4458 (2014).

\bibitem{phosphorene4} S. P. Koenig, R. A. Doganov, H. Schmidt, A. H. Castro Neto, and B. Özyilmaz, Electric field effect in ultrathin black phosphorus, Appl. Phys. Lett. \textbf{104}, 103106 (2014).

\bibitem{phosphorene5} A. Castellanos-Gomez, L. Vicarelli, E. Prada, J. O. Island, K. L. Narasimha-Acharya, S. I. Blanter, D. J. Groenendijk, M. Buscema, G. A. Steele, J. V. Alvarez, H. W. Zandbergen, J. J. Palacios, and H. S. J. van der Zant, Isolation and characterization of few-layer black phosphorus, 2D Mater. \textbf{1}, 025001 (2014).

\bibitem{arsenene1} C. Kamal and Motohiko Ezawa, Arsenene: Two-dimensional buckled and puckered honeycomb arsenic systems, Phys. Rev. B \textbf{91}, 085423 (2015).

\bibitem{arsenene2} M. Pumera, and Z. Sofer, 2D monoelemental arsenene, antimonene, and bismuthene: beyond black phosphorus, Adv. Mater. \textbf{29}, 1605299 (2017).

\bibitem{arsenene3} H. Tsai, S. Wang, C. Hsiao, C. Chen, H. Ouyang, Y. Chueh, H. Kuo, and J. H.  Liang, Direct synthesis and practical bandgap estimation of multilayer arsenene nanoribbons, Chem. Mater. \textbf{28}, 425 (2016).

\bibitem{arsenene4} Z. Zhang, J. Xie, D. Yang, Y. Wang, M. Si and D. Xue, Manifestation of unexpected semiconducting properties in few-layer orthorhombic arsenene,  Appl. Phys. Express \textbf{8}, 055201 (2015).

\bibitem{arsenene5} Y. Wang, Y. Ding, Electronic structure and carrier mobilities of arsenene and antimonene nanoribbons: A first-principle study, Nanoscale Res. Lett. \textbf{10}, 254 (2015).

\bibitem{arsenene6} M. Zeraati, S. M. V. Allaei, I. A. Sarsari, M. Pourfath, and D. Donadio, Highly anisotropic thermal conductivity of arsenene: An ab initio study, Phys. Rev. B \textbf{93}, 085424 (2016).

\bibitem{anisotropic1} K. Dolui and S. Y. Quek, Quantum-confinement and structural anisotropy result in electrically-tunable dirac cone in few-layer black phosphorous, Sci. Rep. \textbf{5}, 11699 (2015).

\bibitem{anisotropic2} S. Das, W. Zhang, M. Demarteau, A. Hoffmann, M. Dubey, and A. Roelofs, Tunable transport gap in phosphorene, Nano Lett. \textbf{14}, 5733 (2014).

\bibitem{anisotropic3} J. Kim, S. S. Baik, S. H. Ryu, Y. Sohn, S. Park, B.-G. Park, J. Denlinger, Y. Yi, H. J. Choi, and K. S. Kim, Observation of tunable band gap and anisotropic Dirac semimetal state in black phosphorus, Science \textbf{349}, 723 (2015).

\bibitem{anisotropic4} S. Yuan, E. van Veen, M. I. Katsnelson, and R. Roldán, Quantum Hall effect and semiconductor-to-semimetal transition in biased black phosphorus, Phys. Rev. B \textbf{93}, 245433 (2016).

\bibitem{anisotropic5} T. Low, A. S. Rodin, A. Carvalho, Y. Jiang, H. Wang, F. Xia, and A. C. Neto, Tunable optical properties of multilayer black phosphorus thin films, Phys. Rev. B \textbf{90}, 075434 (2014).

\bibitem{anisotropic6} T. Low, R. Roldán, H. Wang, F. Xia, P. Avouris, L. M. Moreno, and F. Guinea, Plasmons and Screening in Monolayer and Multilayer Black Phosphorus, Phys. Rev. Lett. \textbf{113}, 106802 (2014).

\bibitem{anisotropic7} T. Low, M. Engel, M. Steiner, and P. Avouris, Origin of photoresponse in black phosphorus phototransistors, Phys. Rev. B \textbf{90}, 081408(R) (2014).

\bibitem{anisotropic8} R. Peng, K. Khaliji, N. Youngblood, R. Grassi, T. Low, and M. Li, Midinfrared electro-optic modulation in few-layer black phosphorus, Nano Lett. \textbf{17}, 6315 (2017).

\bibitem{anisotropic9} K. Khaliji, A. Fallahi, L. Martin-Moreno, and T. Low, Tunable plasmon enhanced birefringence in ribbon array of anisotropic two-dimensional materials, Phys. Rev. B. \textbf{95}, 201401(R) (2017).

\bibitem{SplitOperator} A. Chaves, G. A. Farias, F. M. Peeters, and R. Ferreira. The split-operator technique for the study of spinorial wavepacket dynamics, Commun. Comput. Phys. \textbf{17}, 850 (2015).

\bibitem{SplitOperator0} Kh. Yu. Rakhimov, A. Chaves, G. A. Farias, and F. M. Peeters, Wavepacket scattering of Dirac and Schrödinger particles on potential and magnetic barriers, J. Phys.: Condens. Matter \textbf{23}, 275801 (2011).

\bibitem{SplitOperator1} J. M. Pereira Jr., F. M. Peeters, A. Chaves, and G. A. Farias, Klein tunneling in single and multiple barriers in graphene, Semicond. Sci. Technol. \textbf{25}, 033002 (2010).

\bibitem{SplitOperator3} A. A. Sousa, A. Chaves, T. A. S. Pereira, G. A. Farias, and F. M. Peeters, Quantum tunneling between bent semiconductor nanowires, J. Appl. Phys. \textbf{118}, 174301 (2015).

\bibitem{SplitOperator4} A. Chaves, G. A. Farias, F. M. Peeters, and B. Szafran, Wave packet dynamics in semiconductor quantum rings of finite width, Phys. Rev. B \textbf{80}, 125331 (2009).

\bibitem{SplitOperator6} A. A. Sousa, A. Chaves, G. A. Farias, and F. M. Peeters, Braess paradox at the mesoscopic scale, Phys. Rev. B \textbf{88}, 245417 (2013).

\bibitem{SplitOperator7} A. Chaves, L. Covaci, Kh. Yu. Rakhimov, G. A. Farias, and F. M. Peeters, Wave-packet dynamics and valley filter in strained graphene, Phys. Rev. B \textbf{82}, 205430 (2010).

\bibitem{SplitOperator8} D. R. da Costa, Andrey Chaves, S. H. R. Sena, G. A. Farias, and F. M. Peeters, Valley filtering using electrostatic potentials in bilayer graphene, Phys. Rev. B \textbf{92}, 045417 (2015).

\bibitem{SplitOperator9} D. R. da Costa, A. Chaves, G. A. Farias, L. Covaci, and F. M. Peeters, Wave-packet scattering on graphene edges in the presence of a pseudomagnetic field, Phys. Rev. B \textbf{86}, 115434 (2012).

\bibitem{SplitOperator10} L. S. Cavalcante, A. Chaves, D. R. da Costa, G. A. Farias, and F. M. Peeters, All-strain based valley filter in graphene nanoribbons using snake states, Phys. Rev. B \textbf{94}, 075432 (2016).

\bibitem{SplitOperator11} A. Chaves, D. R. da Costa, G. O. de Sousa, J. M. Pereira Jr., and G. A. Farias, Energy shift and conduction-to-valence band transition mediated by a time-dependent potential barrier in graphene, Phys. Rev. B \textbf{92}, 125441 (2015).

\bibitem{SplitOperator12} D. R. da Costa, A. Chaves, G. A. Farias, and F. M. Peeters, Valley filtering in graphene due to substrate-induced mass potential, J. Phys.: Condens. Matter \textbf{29}, 215502 (2017).

\bibitem{SplitOperator13a} H. M. Abdullah, D. R. da Costa, H. Bahlouli, A. Chaves, F. M. Peeters, and B. Van Duppen, Electron collimation at van der Waals domain walls in bilayer graphene, Phys. Rev. B \textbf{100}, 045137 (2019).

\bibitem{SplitOperator13b} S. M. Cunha, D. R. da Costa, G. O. de Sousa, A. Chaves, J. M. Pereira, G. A. Farias, Wave-packet dynamics in multilayer phosphorene, Phys. Rev. B \textbf{99}(23), 235424 (2019).

\bibitem{SplitOperator13} F. F. Batista Jr. Andrey Chaves, D. R. da Costa, and G. A. Farias, Curvature effects on the electronic and transport properties of semiconductor films, Physica E Low Dimens. Syst. Nanostruct. \textbf{99}, 304 (2018).

\bibitem{ReS2} H.-X. Zhong, S. Gao, J.-J. Shi, and L. Yang, Quasiparticle band gaps, excitonic effects, and anisotropic optical properties of the monolayer distorted $1T$ diamond-chain structures ${\text{ReS}}_{2}$ and ${\text{ReSe}}_{2}$, Phys. Rev. B \textbf{92}, 115438 (2015).

\bibitem{TiS3} J. Dai and X. C. Zeng, Titanium trisulfide monolayer: Theoretical prediction of a new direct‐gap semiconductor with high and anisotropic carrier mobility, Angew. Chem., Int. Ed. \textbf{54}, 7572 (2015).

\bibitem{andreyStarkEffect} A. Chaves, Tony Low, P. Avouris, D. \c{C}akır, and F. M. Peeters, Anisotropic exciton Stark shift in black phosphorus, Phys. Rev. B \textbf{91}, 155311 (2015).

\bibitem{gabriel} G. O. de Sousa, D. R. da Costa, Andrey Chaves, G. A. Farias, and F. M. Peeters, Unusual quantum confined Stark effect and Aharonov-Bohm oscillations in semiconductor quantum rings with anisotropic effective masses, Phys. Rev. B \textbf{95}, 205414 (2017).

\bibitem{edgestate1} H. van Houten, C. W. J. Beenakker, J. G. Williamson, M. E. I. Broekaart, P. H. M. van Loosdrecht, B. J. van Wees, J. E. Mooij, C. T. Foxon, and J. J. Harris, Coherent electron focusing with quantum point contacts in a two-dimensional electron gas, Phys. Rev. B \textbf{39}, 8556 (1989).

\bibitem{edgestate2} G. Montambaux, Semiclassical quantization of skipping orbits, Eur. Phys. J. B \textbf{79}, 215 (2011).

\bibitem{quantumhall1} A. H. MacDonald, Edge states in the fractional-quantum-Hall-effect regime, Phys. Rev. Lett. \textbf{64}, 220 (1990).

\bibitem{quantumhall2} N. Aoki, C. R. da Cunha, R. Akis, D. K. Ferry, and Y. Ochiai, Imaging of integer quantum Hall edge state in a quantum point contact via scanning gate microscopy, Phys. Rev. B \textbf{72}, 155327 (2005).

\bibitem{quantumhall3} M. B\"uttiker, Absence of backscattering in the quantum Hall effect in multiprobe conductors, Phys. Rev. B \textbf{38}, 9375 (1988).

\bibitem{manolopoulos} D. E. Manolopoulos, Derivation and reflection properties of a transmission-free absorbing potential., J. Chem. Phys. \textbf{117}, 9552 (2002).

\bibitem{nonspecular} Y. Betancur-Ocampo, F. Leyvraz, and T. Stegmann, Electron optics in phosphorene pn junctions: Negative reflection and anti-super-Klein tunneling, Nano Lett. \textbf{19}, 7760 (2019).

\end{references}
\end{document}